# THEORETICAL ASPECT OF ENHANCEMENT AND SATURATION IN EMISSION FROM LASER PRODUCED PLASMA

V. N. Rai

Indus Synchrotron Utilization Division

Raja Ramanna Centre for Advanced Technology

Indore-452013 (INDIA)

vnrai@rrcat.gov.in

## ABSTRACT

This paper presents a simplified theoretical model for the study of emission from laser produced plasma to better understand the processes and the factors involved in the onset of saturation in plasma emission as well as in increasing emission due to plasma confinement. This model considers that plasma emission is directly proportional to the square of plasma density, its volume and the fraction of laser pulse absorbed through inverse Bremsstrahlung in the pre-formed plasma plume produced by the initial part of the laser. This shows that plasma density and temperature (that means the electron-ion collision frequency $v_{ei}$) decide the threshold for saturation in emission, which occurs for $v_{ei} \geq 10^{13}$ s$^{-1}$, beyond which plasma shielding effects become dominant. Any decrease in plasma sound (expansion) velocity shows drastic enhancement in emission supporting the results obtained by magnetic as well as spatial confinement of laser produced plasma. The temporal evolution of plasma emission in the absence and presence of plasma confinement along with the effect of laser pulse duration are also discussed in the light of this model.

## 1. INTRODUCTION

Investigation of dynamics and emission from laser-produced plasma under different conditions (Dawson, 1964; Radziemski & Cremers, 1989; Batani, 2010; Schwarz et al., 2010; Kumar et al., 2010a; Kumar et al., 2010b; Kumar et al., 2011; Krasa et al., 2011; Nath & Khare, 2011) is an important subject for many laboratories due to its technological application in various fields of research such as material science, chemical physics, plasma physics as well as inertial and magnetic confinement fusion (Chrisey & Hubler, 1994; Bauerle, 1996; Zel'Dovich & Raizer, 1966; Grun et al., 1981; Radziemski & Cremers, 1989; Rai et al., 1998; Doria et al., 2004; Borisenko et al., 2008; Hoffman, 2009; Wang et al., 2011; Kumar & Verma, 2011; Huber et al., 2011; Fazeli et al., 2011). Properties of laser produced plasma mainly depend on the characteristics of the laser being used for producing the plasma. Various types of phenomena take place in the plasma depending on its density and temperature such as emission of radiation ranging from visible to X-ray wavelength, generation of high energy electrons and ions as well as different waves and instabilities (Radziemski & Cremers, 1989). Any change in the free expansion of plasma affects all of its associated properties. Optical emission spectroscopy of the plasma produced by laser-matter interaction has been investigated by many researchers in the last two decades in connection with laser-induced plasma (LIP) emission as well as laser-induced breakdown spectroscopy (LIBS). LIBS has been established as an analytical tool for in-situ determination of the elemental composition of samples in any form (solid, liquid, gas and aerosols) with fast response and high sensitivity without any sample preparation and surface treatment (Cabalin & Laserna 1998; Babushok et al., 2006; De Giacomo et al., 2007; Singh & Thakur, 2007; Godwal et al., 2008; Schwarz et al., 2010; Nath & Khare, 2011). Along with the wide applicability of LIBS in various fields, it has been used for the characterization of laser-induced plasma processes occurring during the production of thin solid films and nanoparticles of different materials (De Giacomo et al, 2001; Puretzky et al., 2000; Wang et al., 2007; Wang et al., 2011), the control of industrial processes and medical applications (Noll et al., 2001; Sun et al., 2000). These investigations have led to a good understanding of the fundamental processes occurring in various experimental conditions, particularly the distribution of excited state species in the plasma depend both on fluid dynamics of the ablated particles and the balancing of elementary processes (Capitelli et al., 2004; Batani, 2010). These aspect have been described extensively both theoretically and experimentally, and related models have been applied successfully to the interpretation of data obtained in a wide range of conditions (De Giacomo et al., 2004; Capitelli et al., 2004; Mao et al., 2000; Rai et al., 2008; Krasa et al., 2011). This suggests that theoretical study of plasma processes during LIBS can be beneficial in many ways particularly in getting optimum experimental parameters for this technique and enhancing the sensitivity of this analytical system.

Various techniques (Singh and Thakur, 2007; Fazeli et al., 2011; Kumar & Verma, 2011) have been reported to increase the plasma emission (visible to X-rays), which decides the signal and sensitivity of LIBS. Mainly the dual pulse excitation scheme (Babushok et al., 2006; Rai et al., 2003a) as well as magnetic and spatial confinement of the plasma are important techniques (Rai et al., 2003b; Guo et al., 2011), which increase the plasma emission by an order of magnitude. It is supposed that during the plasma confinement kinetic energy of the plasma is transformed into thermal energy, which helps in heating and exciting the atomic species of plasma resulting in enhanced plasma emission. Recently a new technique of plasma confinement has been found that uses the combination of magnetic and spatial confinement (Guo et al., 2011), which effectively improves the emission from the plasma manifold (24 times). In this case, the plasma is confined by a magnetic field as well as by the reflected shock wave from the wall of a small cavity. It has been reported that the combination of both the techniques compresses the plasma at the center of cavity resulting in an increased rate of collisions among the plasma particles, which leads to an increase in the number of atoms in the higher energy state and hence enhanced emission intensity. However the saturation of atomic emission from the plasma inherently limits the laser fluence coupled to plasma that is used for optimizing the detection sensitivity. The saturation phenomenon (plasma emission and material ablation) has been studied by many researchers (Bauerle, 1996; Rai et al., 2003b; Fang & Ahmed, 2007; Batani, 2010) for solids as well as liquid samples, which takes care about dynamic equilibrium between laser energy absorption and plasma expansion as well as consequent cooling. Saturation in plasma emission has also been reported with an increase in laser intensity, which occurs comparatively at lower intensity in the presence of magnetic confinement (Rai et al., 2003c). The plasma shielding effect in laser ablation (Zeng et al, 2005; Batani, 2010 and Leitz et al., 2011) also plays an important role in this process. Therefore the study of the saturation process and its dependence on different plasma parameters is important for enhancing the efficiency of plasma emission. The variation in emission intensities with the focusing distance and the laser pulse energy is related to shielding effects of the plasma produced, which depends on the type of absorption wave obtained at different power densities during the initial plasma formation process (Aguilera et al., 1998). The theoretical basis for enhancement in plasma emission as a result of magnetic confinement as well as under double pulse excitation has been reported earlier (Rai et al., 2003b; Rai et al., 2008). Still very little theoretical work is known about the process of plasma emission and saturation under the effect of variation in plasma parameters and confinement.

This paper presents a simple analytical model to explain the enhancement in emission from laser produced plasma after the confinement (irrespective of confinement techniques). Effect of various other plasma parameters as the plasma density and temperature as well as the laser pulse duration affecting the emission characteristics and saturation is also discussed.

## 2.    THEORY OF PLASMA EMISSION

Vaporization and atomization of a major portion of the sample after laser matter interaction is preceded by heating of samples, which leads to phase changes and ejection of materials away from the surface. Heating of conducting solids by a short laser pulse is essentially thermal in nature. Laser photons are absorbed by electrons in the conduction band of the material, and the energy is given up by collisions with other electrons and lattice phonons, that are converted into heat. The time between collisions is on the order of $\sim 10^{-13}$ s, which is a very short period of time compared to $\sim 10^{-8}$ s of laser pulse duration. In this case, when the incident laser flux density reaches a value in excess of $\sim 10^{8}$ W cm$^{-2}$, the ejected vapor becomes ionized and begins to absorb part of the incident laser energy. At atmospheric pressure, an atmospheric shock wave may precede the ejected material, absorbing a large fraction of the laser energy before the energy reaches the sample material followed by an emission from the plasma (Radziemski & Cremers, 1989). This indicates that plasma formation using single pulse laser can be understood as generated by the combination of two pulses, the initial portion of the laser pulse (Pedestal) and the main portion of the laser pulse. For this it is necessary to understand first the case of dual laser pulse excitation followed by emission from plasma and then to simplify it for emission from single pulse laser produced plasma.

## 2.1    Plasma Emission from Dual Pulse Laser

Normally double pulse LIBS (Singh and Thakur, 2007) uses two laser pulses separated by microsecond time delay. The first laser creates the plasma expanding normal to the target surface, whereas the second laser pulse gets absorbed in it followed by further ablation of target material by the residual second laser pulse. The absorption of second laser in the plasma created by the first laser pulse can be assumed to be dominated by inverse Bremsstrahlung absorption, which occurs due to electron – ion collisions in the plasma. If plasma electrons are subjected to momentum changing collisions as they oscillate back and forth in the laser electric field, the laser light wave undergoes an effective damping. The spatial damping rate of wave energy, $k_{ib}$ can be given by (Max, 1982)

$$k_{ib} \propto \frac{Z n_e^2}{T_e^{3/2} \left(1 - \frac{n_e}{n_c}\right)^{1/2}} \tag{1}$$

Where $n_e$ and $T_e$ are the density and temperature of the plasma and Z is the atomic number of the target material. This shows that inverse Bremsstrahlung is strongest for the low temperature, high density and high Z target materials. According to eq. (1) a uniform plasma of length L and density $n_e < n_c$ will produce a one way absorption fraction given by

$$\alpha_{abs} = 1 - \exp\left(-k_{ib} L\right) \tag{2}$$

For weak absorption $k_{ib}L << 1$ and $\alpha_{abs} \cong k_{ib}L$. For strong absorption, $k_{ib}L >> 1$ and $\alpha_{abs} \rightarrow 1$. This indicates that absorption fraction is linear in $k_{ib}L$ for small absorption and then saturates when $k_{ib}L$ is large. Even the long plasma length enhances the absorption. The result of inverse Bremsstrahlung absorption in inhomogeneous plasma is more complicated. A simplified version of absorption fraction for a linear density profile is given by (Max, 1982)

$$\alpha_{abs} = 1 - \exp\left\{-\frac{32}{15}\frac{L\nu_{ei}(L)}{c}\right\} \qquad (3)$$

where $\nu_{ei}$ (L) is the electron-ion collision frequency at critical density and c is the speed of light. This indicates that inverse Bremsstrahlung absorption is strongest for low temperature, high densities, high Z targets and long density scale lengths. It is sensitive to details of the density profile near the critical surface and can be diminished considerably if the profile falls steeply at the critical surface. For simplification purpose we have considered plasma expansion length $L \approx C_sT$, where $C_s$ is the ion acoustic (plasma expansion) speed and T is the plasma expansion time. An expression for the electron–ion collision frequency $\nu_{ei}$ can be given as (Bittencourt, 1986)

$$\nu_{ei} = 3.62 \times 10^{-6} n_i\ T_e^{-3/2} \ln\Lambda\ \ s^{-1} \qquad (4)$$

where $n_i$ ($m^{-3}$) is the ion density in the plasma, $T_e$ (K) is the plasma temperature and $\ln\Lambda$ ~ 10. Electron–ion collision frequency $\nu_{ei}$ is considered to be the constant over the whole plasma length.

In the planar geometry, the volume of the radiating plasma can be approximately given by $V = \pi R^2 L$ for a plasma column of diameter 2R and length L. The mass ablated by the first and the second laser pulse can be given as $\dot{m}_1\pi R^2\tau_L$ and $\dot{m}_2\pi R^2\tau_L$ respectively, where $\dot{m}_1$ and $\dot{m}_2$ are the mass ablation rate after the first and the second laser pulse and $\tau_L$ is the time duration of both the lasers. The parameters of both the lasers are considered to be the same. It is well known that the plasma emission takes place through three processes, the Bremsstrahlung emission, the free-bound transition and the bound-bound transition. In each case, the plasma emission is strongly dependent on the product of electron and ion density ($\propto n_e n_i$), the temperature of plasma as well as on the volume of plasma. Considering that the plasma emission is proportional to the square of the plasma density ($n_e = n_i$) and the plasma volume, one can write an expression for the emission intensity after the first laser pulse ($I_1$) as

$$I_1\ \ \propto\ \ \left(\frac{\dot{m}_1\pi R^2\tau_L}{\pi R^2 C_{s1}T_1}\right)^2\left(\pi R^2 C_{s1}T_1\right) \qquad (5)$$

Where the first factor is the plasma density and the second factor is the volume of the plasma. $C_{s1}$ is the ion acoustic speed of the plasma due to first laser pulse and $T_1$ is the time of plasma expansion that is gate delay from the first laser when the emission was recorded in single pulse LIBS.

The emission after the second laser pulse ($I_2$) can be written as

$$I_2 \propto \left( \frac{\dot{m}_2 \pi R^2 \tau_L}{\pi R^2 C_{s1} \Delta t + \pi R^2 C_{s2} T_2} \right)^2 \left( \pi R^2 C_{s1} \Delta t + \pi R^2 C_{s2} T_2 \right) \left[ 1 - \exp\left\{ -\frac{32}{15} \frac{C_{s1} \Delta t \, \nu_{ei}}{c} \right\} \right] \qquad (6)$$

Here the first factor describes the plasma density after the time ($\Delta t + T_2$), the second factor represents the plasma volume after the time ($\Delta t + T_2$) and the third factor equals the fraction of the second laser absorbed in the plume of the plasma created by the first laser pulse. $\Delta t$ is the time delay between the lasers, $C_{s2}$ describes the ion acoustic speed after the second laser pulse and $T_2$ is the time (gate delay) after firing the second laser. To deduce a simplified expression for enhancement in double pulse emission and canceling all the common factors affecting the single and double pulse LIBS emission, the enhancement (E) is obtained by normalizing the sum of plasma emission under single and dual pulse LIBS by the emission from single pulse LIBS (Rai et al., 2008).

$$E = \frac{I_1 + I_2}{I_1} = 1 + \frac{I_2}{I_1}$$

$$E = 1 + \left( \frac{\dot{m}_2}{\dot{m}_1} \right)^2 \frac{\left[ 1 - \exp\left\{ -\frac{32}{15} \frac{C_{s1} \Delta t \, \nu_{ei}}{c} \right\} \right]}{\left( \frac{\Delta t}{T_1} + \frac{C_{s2}}{C_{s1}} \frac{T_2}{T_1} \right)} \qquad (7)$$

where one can take for the value of $T_1$ any value between $\Delta t$ (for maximum emission) and the time of peak emission in double pulse LIBS (from first laser pulse) without any significant change in the result. Eq. (7) has been used to study the effect of different parameters on the enhancement in plasma emission under dual pulse excitation, which explained various observations of dual pulse LIBS as reported earlier (Rai et al., 2008) . According to eq. (7) zero delay between lasers ($\Delta t = 0$) provides an enhancement of unity (E =1), just like a single pulse LIBS emission.

## 2.2    Plasma Emission from Single Pulse laser

The above eq.-7 obtained for dual pulse excitation of the plasma can be simplified for single pulse case under the light of above discussion that the plasma is formed and starts expanding when the initial portion of laser pulse interacts with the matter. So a single laser pulse can also be considered operating as dual pulse with certain assumptions. Mainly one can consider that after the plasma formation by pedestal of the laser, it expands for the period $\tau_L$ (duration of laser pulse) during which the laser gets absorbed in the expanding plasma. This indicates that $\Delta t$ delay between two lasers in eq.-7 can be replaced by $\tau_L$ for the single pulse case. The eq.-7 can be simplified in two ways as described below to better understand the dependence of important plasma parameters on the emission characteristics of the plasma as well as on its temporal behavior.

**Case - 1**

In this case assumption is made that the $\Delta t = \tau_L$ (laser time duration), $T_1 = \tau_L/2$ (time of peak laser intensity). $T_2$ is the time at which the measurement of the plasma emission is taking place. Other consideration is that $C_{s2} = 2\,C_{s1}$. In this situation eq.-7 can be written as

$$E = 1 + \cfrac{\left(\dfrac{\dot{m}_2}{\dot{m}_1}\right)^2 \left[1 - \exp\left\{-\dfrac{32}{15}\dfrac{C_{s1}\tau_L V_{ei}}{c}\right\}\right]}{\left(2 + 4\dfrac{T_2}{\tau_L}\right)} \qquad (8)$$

This equation can provide information about the emission from the plasma at different time duration that means the variation of emission with the gate delay ($T_2$).

**Case – 2**

In this case we consider $\Delta t = \tau_L$ and $T_1 = T_2 = T$ is the gate delay or the time of measurement of the emission after the laser peak. Ratio of the laser pulse time duration $\tau_L$ and gate delay T (time of measurement) will be negligibly small for nanosecond time duration laser ($\tau_L / T \approx 0$). After these assumption eq.-7 can be written as

$$E = 1 + \left(\frac{\dot{m}_2}{\dot{m}_1}\right)^2 \left(\frac{C_{s1}}{C_{s2}}\right)\left[1 - \exp\left\{-\frac{32}{15}\frac{C_{s1}\tau_L \nu_{ei}}{c}\right\}\right] \qquad (9)$$

This indicates that the plasma emission under single pulse laser plasma interaction is mainly dependent on the square of the ratio of the mass ablation rate after full pulse and initial mass ablation, the ratio of plasma sound velocity, the plasma collision frequency and the pulse duration of laser.

## 3.      RESULTS AND DISCUSSION

In order to achieve a better understanding of the physical processes arising in single pulse LIBS, we have estimated the intensity of emission for different plasma parameters arising during the laser plasma interaction. The important parameters are: the mass ablation rate, plasma density, plasma temperature, electron-ion collision frequency as well as the laser time duration. In fact, the effect of laser intensity on the plasma emission will be taken care by the increase in plasma density and temperature as both are directly proportional to the laser intensity.

### 3.1      Effect of Material Ablation

Fig.-1 shows the variation in the emission intensity calculated theoretically using eq.-9 for different values of the ratio of mass ablation rate of full laser pulse and its pedestal given as $\frac{\dot{m}_2}{\dot{m}_1}$. For this calculation $\nu_{ei} \sim 3.62 \times 10^{12}$ s$^{-1}$ is taken corresponding to a plasma density of $10^{17}$ cm$^{-3}$ and a plasma temperature of 1 eV. The ratio of plasma sound velocity of full intensity plasma and initial (pedestal) plasma is kept as $\frac{C_{s2}}{C_{s1}} \sim 2$. The variation has been shown for 1, 10 and 20 ns laser pulses. The variation shows that the emission intensity increases as the value of $\frac{\dot{m}_2}{\dot{m}_1}$ increases. Initially increase in emission intensity is slow but for the higher value of ratio of mass ablation rates, it increases fast as emission is proportional to the square of ratio of mass ablation rates. Normally material ablation is directly related with the laser intensity, which indicates that ultimately an increase in the laser intensity increases the plasma emission. The time duration of the laser pulse also plays an important role in plasma emission as emission intensity increases with an increase in the time duration of laser. These observations are as expected because an increase in interaction time keeps the plasma hot for a longer time as a result of efficient absorption of the laser providing more emission from the plasma. These results are in agreement with the experimental observations reported earlier (Zeng et al., 2005; Leitz et al., 2011).

It is well known that many parameters decide about the initial ablation of sample material by pedestal of the laser at very low intensity (Piepmeier, 1986). These parameters, such as sample reflectivity, thermal conductivity, vaporization temperature, heat capacity and latent heats of fusion and vaporization, are important in determining the total amount of material removed from the sample. At starting of the laser pulse, electrons and ions are released from the surface by photoelectric and thermionic processes. The free electrons above the surface gain energy by absorbing photons from the laser beam. This initial plasma may trigger the generation of more plasma that is formed by thermal ionization of the vaporizing bulk material of the sample. Initially at lower irradiances, the sample material at the surface is not heated much above its boiling temperatures. In this case velocity of the vaporizing materials is on the order of $\sim 10^4$ cm $s^{-1}$. However at higher irradiances the velocities of the sample species reach values above $10^6$ cm $s^{-1}$ corresponding to the particles with energies above their normal boiling temperatures (Piepmeier, 1986). In this situation X-rays and highly ionized species are also observed indicating that energies of the plasma reached from ten to hundreds of electron volts, which comes down after expansion. Individual particles of high energy result from the highly energetic plasma generated by inverse Bremsstrahlung absorption of the high power density of laser beam. The amount of material ablated depends on the time duration of the laser. It is several orders of magnitude lesser for short time duration laser (Piepmeier, 1986, Zeng et al., 2005; Leitz et al., 2011). This is because much shorter time is available for conduction of heat into the sample. This indicates that energy of the shorter laser pulse is therefore useful in processes near the surface rather than in vaporizing material at greater depth. Crater depth in material ranges from 1 to 10 µm for a Q switched laser and the amount of material ablated is generally at and below the microgram level (Piepmeier 1986). A comparison shows that short pulse micro and nanosecond laser systems generally allow high ablation rates, where the possibility of thermal damage of the target piece is higher. On the other hand ultrafast picosecond and femtosecond laser systems provide lower ablation efficiency but allow a higher precision in ablation (Leitz et al., 2011).

## 3.2    Effect of Plasma Density (Laser Energy)

It is well known that during laser matter interaction an increase in laser intensity increases the plasma density as well as plasma temperature. Plasma density increases due to increased material ablation, whereas plasma temperature increases due to absorption of laser energy into plasma preformed due to pedestal (initial part) of the laser. To better understand the effect of plasma density and temperature on the process of plasma emission, calculation was made using eq.-9 for different values of electron-ion collision frequencies, because it is decided by the value of plasma density and temperature. Ultimately this collision frequency plays an important role in absorption of laser energy into plasma. For this calculation we took $\frac{\dot{m}_2}{\dot{m}_1} = 4$ and $\frac{C_{s2}}{C_{s1}} = 2$. Fig.-2 shows the variation in emission intensity with an increase in plasma density ($10^{13}$ to $10^{20}$ cm$^{-3}$) for different values of plasma temperature (1, 5 and 10 eV). It shows that initially plasma emission increases slowly followed by a fast nearly linear increase in emission between $10^{15}$ to $10^{19}$ cm$^{-3}$. However, a further increase in plasma density beyond $10^{19}$ cm$^{-3}$ shows saturation in emission. It has been noted that the emission intensity is high for lower value of plasma temperature, which is expected as the value of electron-ion collision frequency remains high for low plasma temperature. This increases the absorption of the laser intensity by the plasma, which results in a stronger emission. Another observation is that emission intensity starts increasing and saturating at comparatively lower value of plasma density for low temperature plasma. Again the role of plasma frequency is found to be more important. Finally one can say that an increase in laser intensity increases the plasma density as well as the temperature, where the balance between the two effects leads towards saturation in emission.

It is important to compare the results of this calculation with the experimental observations. Rai et al. (2003b) has reported the effect of laser energy on the LIBS signal in the presence of magnetic field. LIBS spectra of aluminum alloy sample were recorded in the absence as well as in the presence of a magnetic field, where the energy of the laser was varied from 10-150 mJ ($4 \times 10^9$-$60 \times 10^9$ W/cm$^2$). The variation of chromium line intensity ($\lambda$=357.87 nm) with laser energy in the absence of a magnetic field shows that emission intensity increases first very slowly and then very fast up to nearly 20 mJ. Further increase in laser energy saturates the chromium line emission intensity. It seems that initially the laser radiation is absorbed up to 20 mJ, where ablation of the material and consequently density of plasma increases with laser intensity and gets saturated. These observations are similar as obtained by theoretical calculations (Fig.-1&2). However the excess laser energy beyond 20 mJ ($\geq 8 \times 10^9$ W/cm$^2$) is not being coupled (absorbed) to laser produced plasma. This is possible only due to shielding produced by the plasma near critical density ($10^{21}$ cm$^{-3}$) surface formed at high laser intensity where electron-ion collision frequency is $> 10^{13}$ s$^{-1}$. Similar behavior of plasma emission has been reported (Aguilera et al., 1998). Probably excess laser energy is being reflected from the critical density surface of the plasma, where the plasma frequency becomes equal to the laser frequency. In this case, laser can not propagate toward the higher plasma density side. At higher plasma density, self-absorption of emission is also possible leading to saturation, when the experiment is performed at higher laser intensity. The presence of self-absorption can be observed from the measurement of full width at half maximum (FWHM) of the Lorenzian profile of line emission (Aguilera et al., 1998). Both the processes are possible in the high density plasma, which can contribute towards saturation (or decrease in saturation) in the emission. The breakdown in gases surrounding the target also shields the laser energy reaching the target surface, which results in decrease and/or saturation in material ablation from the target surface and consequently in the plasma emission (Batani, 2010; Lee et al., 2011). Rai et al. (2003b) further report that the presence of magnetic field shows no significant effect on atomic line emission intensity, when the laser energy was below 10 mJ (when the plasma density is low). However an increase in the emission intensity in the presence of a magnetic field was noted for laser energy between 10 to 50 mJ. This enhancement in emission in the presence of magnetic field is due to an increase in the plasma density (electron-ion collision frequency) as a result of magnetic confinement of the plasma, which is effective at lower plasma temperature, when the laser intensity is lower. However, this enhancement in emission intensity decreased, when the laser energy was further increased beyond 50 mJ. The plasma emission in the presence of a magnetic field was even lower than the emission observed in the absence of a magnetic field for laser energy > 80 mJ ($32\times10^9$ W/cm$^2$). It seems that along with reflection of laser energy from critical density level and self-absorption of emission, some other loss factors are also added up in the presence of magnetic field. This extra factor may be due to the presence of density fluctuation (instability) in the plasma, which is generated at higher laser intensity in the presence of a magnetic field. Density fluctuations are expected in this situation due to bouncing of the plasma near $\beta = 1$ location, where plasma kinetic energy becomes equal to the magnetic energy. Even some high-frequency instability, such as large Larmor radius instability are also expected in such plasma condition (Cabalin & Laserna, 1998; Rai et al., 2003b). The presence of instability in the plasma can lead to the scattering of laser light as well as the loss of plasma particles as a result of cross field diffusion. This may be the reason behind the decrease in emission intensity even after saturation, when the laser energy is more than 80 mJ in the presence of magnetic field. Similar behaviour is noted for iron line emission at $\lambda = 358.12$ nm with an increase in laser energy in the absence and presence of magnetic field.

In another plasma experiment (Rai et al., 2003c) the plasma is formed from liquid jet target having magnesium (Mn) in low concentration. It is noted that the intensity of all the three Mn lines increases linearly with laser energy up to ~ 300 mJ in the absence of magnetic field. But the presence of the magnetic field shows an enhancement in the LIBS intensity up to 200 mJ, followed by a saturation/ decrease towards higher laser energy. The maximum signal enhancement was observed between 150 - 200 mJ of laser energy. Similar enhancement in the emission intensity was noted for other elements also such as Magnesium, chromium and titanium in aqueous solution in the presence of magnetic field. Although the threshold laser energy for the breakdown was ~10 mJ for solid sample in comparison to ~ 50 mJ for a liquid sample, even then the basic observations are similar for solid or liquid samples. The enhancement in the emission intensity was noted mainly for moderate laser energy after the breakdown. Finally these experimental observations are found in agreement with the theoretical calculations.

### 3.3    Saturation in Plasma Emission

The study of variations in plasma emission with electron density and temperature for laser of different time duration can provide better understanding of saturation in plasma emission. For this purpose the calculation is performed using eq.-9 with a consideration that $\frac{\dot{m}_2}{\dot{m}_1} = 4$,    $\frac{C_{s2}}{C_{s1}} = 2$ and $T_e$ =1eV and the results are presented in Fig.-3, which shows the variation in plasma emission with change in plasma density for different time duration of the laser $\tau_L$. It is shown that the plasma emission starts increasing at ~ $10^{16}$ cm$^{-3}$ for a laser of 1 ns time duration. This increase became fast with an increase in plasma density, which saturates at ~ $10^{19}$ cm$^{-3}$. Similar variations are  noted for 10 and 20 ns time duration laser, but with minor change in the threshold for initiation and saturation of emission, which occur at lower density for large time duration lasers. Higher time duration laser produce comparatively more plasma emission as a result of more material ablation (Zeng et al.,2005; Leitz et al., 2011). Here again the electron-ion collision frequency and laser plasma interaction time play an important role. In this case, the plasma collision frequency increases with an increase in the plasma density, which helps in increasing the absorption of laser in plasma through inverse Bremsstrahlung and consequently the plasma emission, whereas large time duration of laser provides longer time for plasma to expand and absorb the laser efficiently.

The effect of plasma temperature on its emission has also been obtained using eq.-9, where the plasma density was taken constant at~$10^{17}$ cm$^{-3}$ keeping all other parameters the same as in Fig.-3. In this case emission from plasma is calculated by varying the plasma temperature for different time duration laser. Fig.-4 shows that plasma emission is very high for low temperature plasma, which decreases first very fast and then slowly with an increase in plasma temperature. The reason for decrease in plasma emission with an increase in plasma temperature is due to decrease in electron-ion collision frequency as

$\nu_{ei} \propto (T_e)^{-3/2}$. Here again plasma emission is high for longer time duration laser as seen in earlier case. In this case saturation is seen at lower plasma temperature near 1 eV. Results of Fig.- 3 & 4 indicates that saturation in the plasma emission is possible only when the plasma density is high and plasma temperature is low. In other words one can say that instead of plasma density and plasma temperature, electron-ion collision frequency play an important role in deciding the saturation in plasma emission, which is $\nu_{ei} > 10^{13}$ s$^{-1}$ as has been discussed earlier.

### 3.4  Laser Absorption and Shielding Effect in Plasma

The observation of  fig.-1 & 2  can be better understood by considering variation in the laser absorption mechanism with an increase in laser intensity. For lower laser intensity a laser-supported detonation wave (LSDW) is expected in which the laser absorption region propagates as a shock wave (Aguilera et al., 1998). In this regime if the electron density remains below the critical density then an increase in laser intensity will increase its absorption in the plasma resulting in more ablation as well as plasma emission. This is the reason why plasma emission increases linearly with laser intensity. When the plasma density reaches critical density level, the plasma becomes opaque to the laser beam, which shields the target. In this case coupling of laser intensity to plasma decreases, which leads to a saturation in plasma emission. This phenomenon can be observed in both types of targets (solid and liquid) when the laser intensity is in the medium range. This saturation takes place at comparatively low laser intensity for solid in comparison to  liquid sample due to production of higher plasma density in the case of solid targets. At higher laser intensity, the absorption mechanism changes to laser-supported radiation wave,  in which plasma temperature becomes very high resulting in a transparent plasma for the laser beam, which starts emitting Bremsstrahlung (background) emission due to its high temperature (Aguilera et al., 1998). The plasma shielding effect is then reduced and the ablation of the target material enhanced. In this regime also plasma emission (x-rays and visible) intensity increases linearly. A saturation in the x-ray emission was also noticed in the form of change in the slope with an increase in the laser intensity. However this saturation was found  to be correlated with the loss of plasma energy in form of the generation of an instability and high energy particles in the plasma along with emission of high energy x-rays (Rai et al., 2000a, Rai et al., 2000b). Normally, the high-intensity laser experiment involves measurement of x-ray emission under vacuum conditions. However, the high laser intensity experiment in air also shows saturation, where the shielding of the laser occurs by the air plasma (Aguilera et al., 1998; Lee et al., 2011). In this case, presence of the target surface enhances breakdown in the surrounding gas, which is why air plasma is also formed when the laser is focused below or on the sample surface (Multari et al., 1996).

## 4.    EFFECT OF PLASMA CONFINEMENT

It is well known that laser produced plasma expands away very fast from the target surface and emits various types of emission ranging from X-ray to visible emission. Both types of plasma emission increase after plasma confinement. Two types of plasma confinement have been reported as magnetic and spatial confinement for enhancing the emission from laser produced plasma.. In both the cases plasma expansion is decelerated. Here we have tried to find the effect of expansion velocity on the plasma emission using eq.-9. For this calculation we have taken $v_{ei} \sim 3.62 \times 10^{12}$ $s^{-1}$ corresponding to a plasma density of $10^{17}$ $cm^{-3}$ and a plasma temperature of 1eV. The value of $C_{s2}/C_{s1}$ has been changed from 0.1 to 10. Fig.-5 shows that the emission intensity is very high when the value of $C_{s2}/C_{s1}$ is low towards 0.1, where it decreases very fast as the value of $C_{s2}$ increases. This observation shows that deceleration in plasma expansion enhances the plasma emission. It seems that the plasma volume increases with an increase in $C_{s2}$ and the corresponding energy density of absorbed laser decreases, which is reflected as decreased plasma emission. This indicates that enhancement in emission is mainly dependent on the order of plasma confinement. This supports the results of enhancement in emission by magnetic and spatial confinement. More enhancement in spatial confinement is due to comparatively better confinement/deceleration of plasma. In the case of magnetic confinement, the plasma diffuses across the magnetic field as a result of plasma instability and fluctuations in the plasma, which degrades the plasma confinement, whereas spatial confinement is free from these problems. However, the combination of both the confinement techniques is beneficial as has been reported by Guo et al., (2011). Enhancement in plasma emission is independent of the technique used for plasma confinement. Only one factor is important, and that is the efficiency of the plasma confinement. This result further indicates that spatial confinement of dual pulse produced plasma may provide even more enhancement in emission than the combination of magnetic and spatial confinement.

## 4.1 Temporal Evolution of Emission from Confined Plasma

The temporal evolution of plasma emission is obtained using eq.-8. For this purpose we have considered $T_1 = \tau_L/2$ (Peak time of laser) and $T_2$ is varied, which is equivalent to the time delay at which the emission intensity is recorded. To see the effect of plasma confinement and the laser time duration on the temporal evolution, calculations have been performed first for different values of $C_{s2}/C_{s1}$ as 2, 0.5 and 0.1 keeping laser time duration constant at 10 ns and then for 1, 10 and 20 ns time duration laser keeping $C_{s2}/C_{s1}$ constant at ~ 2. Fig.-6 has the variation of plasma emission with time delay for decreasing value of $C_{s2}/C_{s1}$, which shows plasma confinement effect. It shows that for $C_{s2}/C_{s1}$ ~ 2 plasma emission first decreases very fast and then slowlywith time delay. After adding the confinement effect by decreasing the values of $C_{s2}/C_{s1}$ to 0.5 and 0.1, plasma emission increases for all the delay time. Even the rate of decay in emission intensity changes, which indicates that plasma emission last for a longer time in the case of plasma confinement. It further shows that maximum enhancement in the emission after confinement occurs around the time delay of 20 ns. This happens because of slowing down of decay in plasma emission as a result of plasma confinement. Fig.-7 shows the variation in temporal evolution of plasma emission obtained from a laser having a time duration of 1, 10 and 20 ns. Here the plasma emission is less for 1 ns laser produced plasma, which initially decay fast and then slowly. The rate of plasma decay decreases with an increase in time duration of the laser. Even the emission is higher for the plasma produced by a longer time duration laser, which is due to efficient heating of the plasma for a longer time as discussed earlier. It also shows that a plasma produced by a long time duration laser lasts for a longer time. These observations are in agreement with the experimental observations (Zeng et al., 2005; Leitz et al., 2011).

Similar observation has been reported experimentally (Rai et al., 2003b). They have studied temporal evolution of chromium line emission from aluminum alloy in the absence and presence of a magnetic field (magnetic confinement) and reported that plasma emission was maximum at ~ 2 µs after which it started decreasing. The emission intensity lasted for nearly 20 μs. An overall increase in intensity was noted for all the delay in the presence of magnetic field (magnetic confinement), whereas the shape of the curve remained the same. This result is similar as shown in Fig.-6. The decay in emission with time indicates that initially the plasma is hot and the emission from it is dominated by the background emission, which contains mainly Bremsstrahlung emission. As the gate delay increases, the plasma expands away from the target and gets cool resulting a decrease in emission. The presence of the magnetic field increases the effective density of plasma in the emission volume as a result of confinement of laser produced plasma, which ultimately increases the probability of radiative recombination, electron-ion collision frequency as well as absorbed energy per unit volume. All of these factors play role in increasing the plasma emission.

During magnetic confinement enhancement in plasma emission and deceleration in plasma expansion changes with plasma β (Rai et al., 2003b). It is well known that plasma β depends mainly on the density and temperature of the plasma, which decreases as the plasma expands away from the target. It also depends on the intensity of magnetic field, which remains constant during this experiment. It is shown that initially for a higher value of β, the plasma deceleration and enhancement in emission is negligible. However as the β decreases below 10, the ratio of plasma expansion velocities $v_2/v_1$ decreases and the ratio of emission intensity $I_2/I_1$ (value of enhancement) increases. This indicates that Bremsstrahlung emission, which occurs at higher plasma temperatures and density (β > 10) and comparatively at early time (smaller gate delay) after the plasma formation, cannot contribute to the enhancement in the emission. Mainly line emissions are getting affected due to recombination radiations or heating of plasma as a result of its confinement. These reported results are also in qualitative agreement with the theoretical findings of this paper.

## 5.  CONCLUSIONS

A simplified model is presented to study the emission from laser produced plasma, which explains qualitatively some of the important results obtained in LIBS and laser plasma interaction experiments. This shows that the saturation in emission is found to be dependent on the time duration of the laser as well as on the plasma density and plasma temperature, particularly on the electron-ion collision frequency. Onset of saturation in the emission becomes effective for electron-ion collision frequency $v_{ei} \geq 10^{13}$ $s^{-1}$. However, the confinement of plasma, which is correlated with the decrease in plasma expansion velocity, increases the plasma emission drastically. The increase in plasma emission is found  to be mainly dependent on the efficiency of the plasma confinement, which can be achieved by any technique. Another observation is that plasma emission decays (decreases) very fast and then slowly with time. However confinement of the plasma makes the decay of the plasma emission slow, due to which the plasma emission lasts for a longer time. These results indicate that the combination of the dual pulse experiment with spatial and magnetic confinement may enhance the plasma emission even more than what is reported till date (Singh & Thakur, 2007; Guo et al., 2011). Similar results are possible even for the X-ray emission from laser produced plasma after its confinement. The results of this study have been obtained after many assumptions, where the inter-dependence of different parameters is not taken into account. However, further investigation is needed for improvement of this model considering various other parameters and its inter-dependence along with an experiment for an efficient confinement of the plasma, which will be beneficial for increasing the plasma emission (visible to X-ray) and consequently the sensitivity of LIBS.

**Acknowledgment**

Author is thankful to Dr. S. K. Deb, Head,  Indus Synchrotron Utilization Division, RRCAT for his encouragement and support.

**FIGURE CAPTION**

1. Variation in emission intensity from plasma with change in ratio of mass ablation rate ($\frac{\dot{m}_2}{\dot{m}_1}$) for plasma produced by 1, 10 and 20 ns time duration laser.

2. Variation in emission intensity with an increase in plasma density for plasma temperature of 1, 5 and 10 eV.

3. Variation in emission intensity with an increase in plasma density for plasma produced by 1, 10 and 20 ns time duration laser.

4. Variation in emission intensity with an increase in plasma temperature for plasma produced by 1, 10 and 20 ns time duration laser.

5. Variation in emission intensity with an increase in ratio of expansion velocity ($C_{s2}/C_{s1}$) for plasma produced by 1, 10 and 20 ns time duration laser.

6. Variation in emission intensity with time of measurement after laser peak for plasma having ratio of expansion velocity ($C_{s2}/C_{s1}$) as 0.1, 0.5 and 2.

7. Variation in emission intensity with time of measurement after laser peak for plasma produced by a laser of 1, 5 and 10 ns time duration.

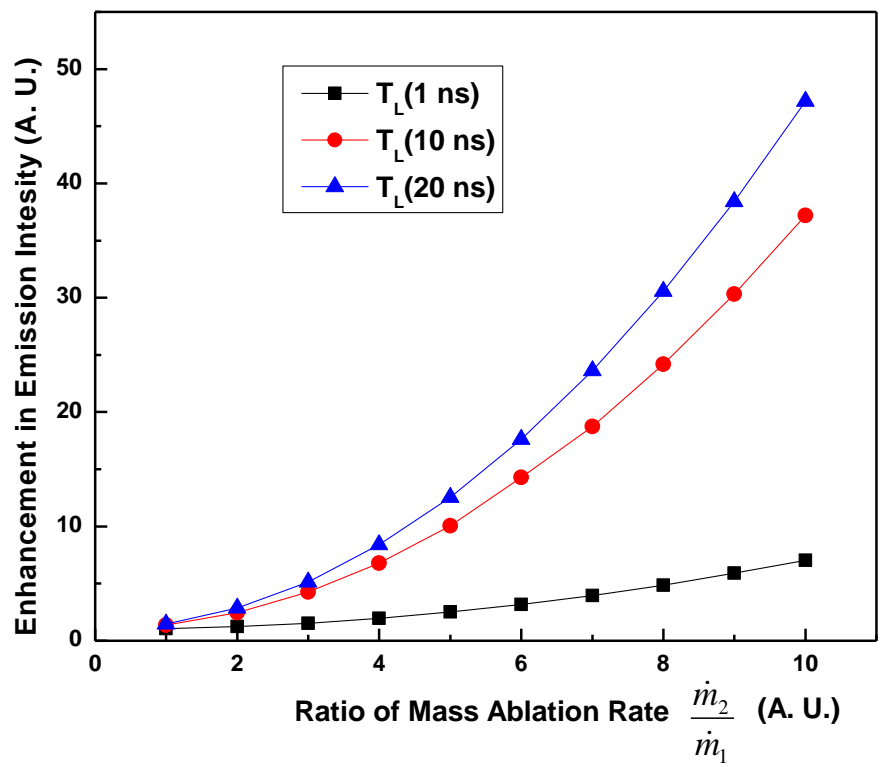

**Fig. - 1**

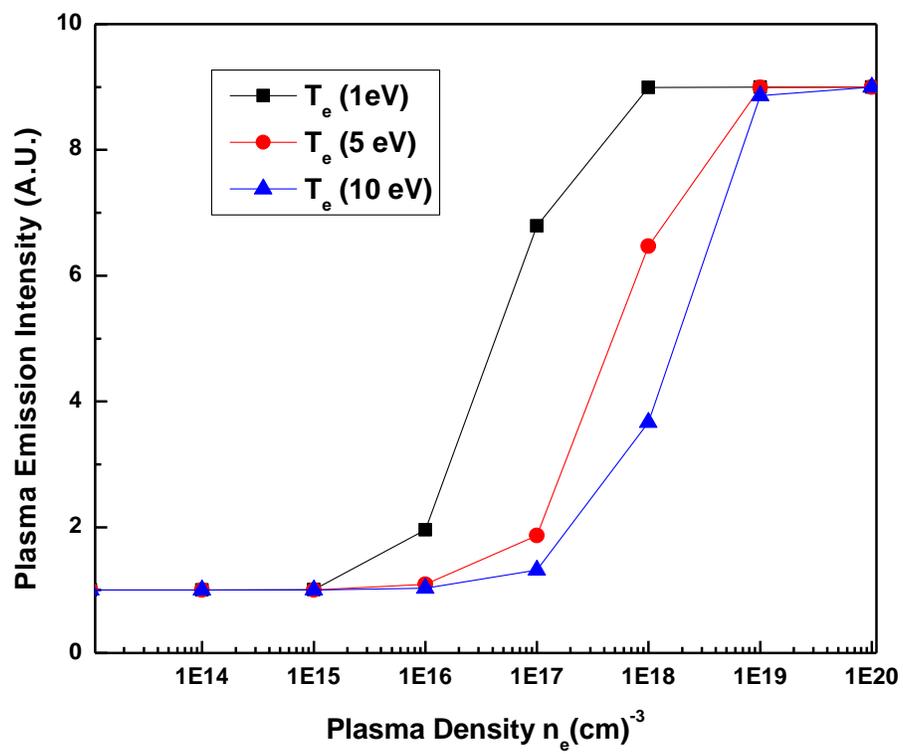

**Fig. - 2**

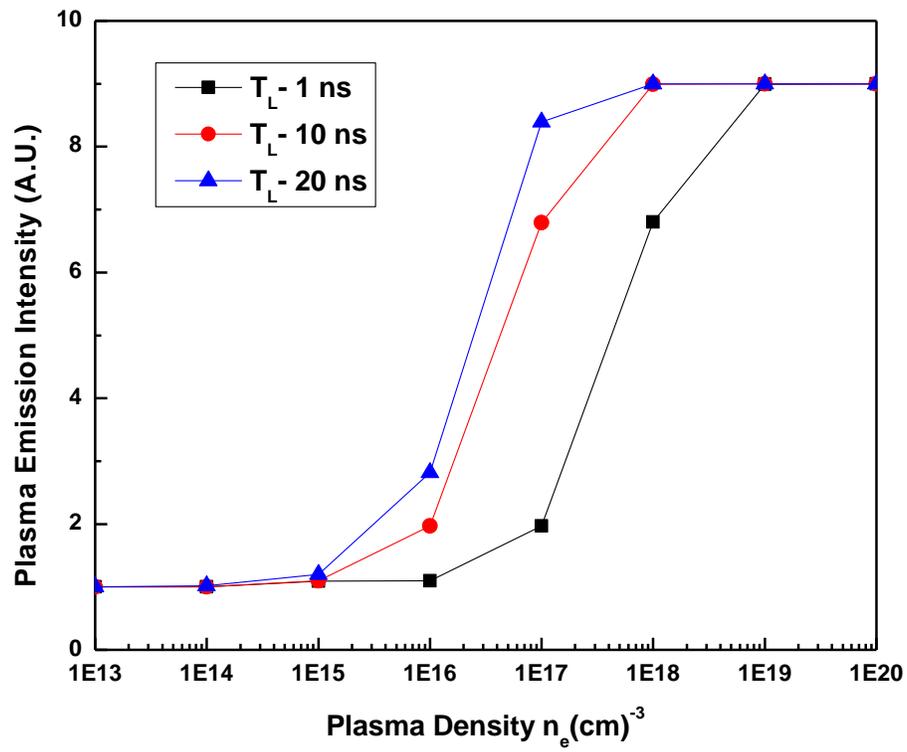

**Fig. - 3**

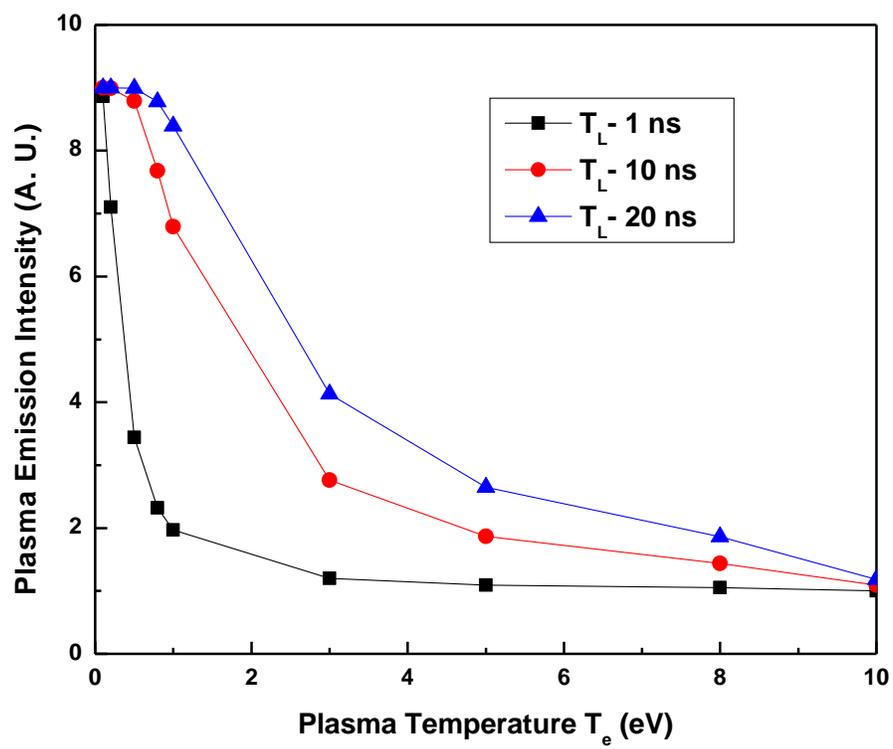

**Fig. - 4**

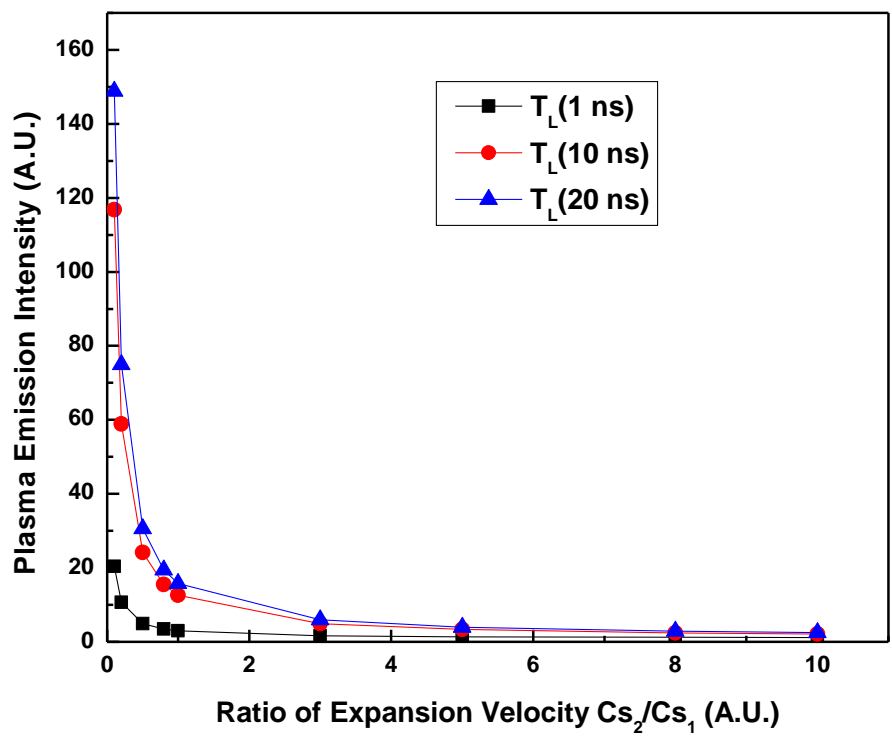

**Fig. – 5**

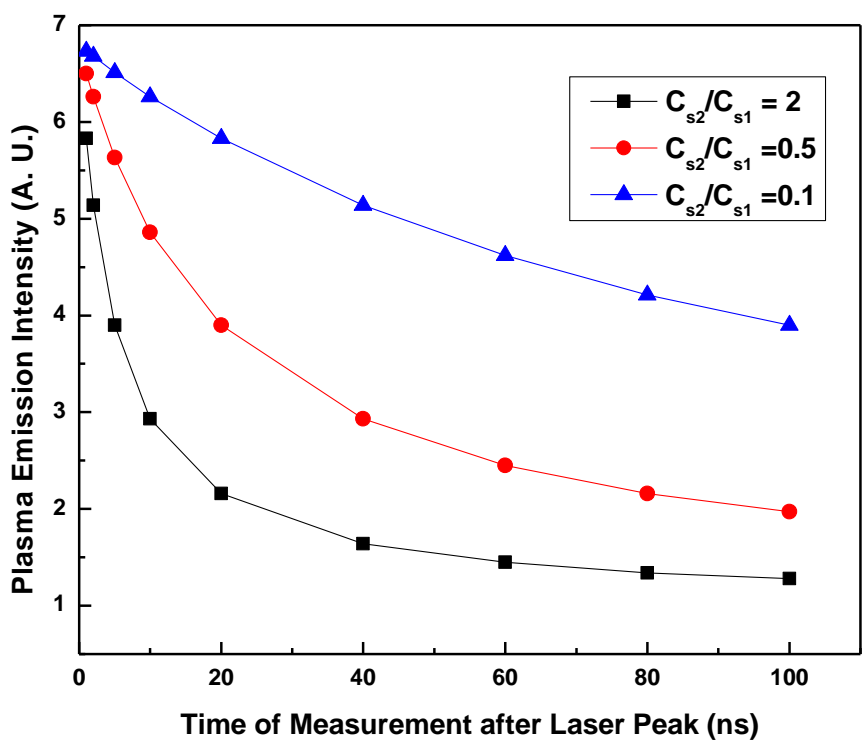

**Fig. – 6**

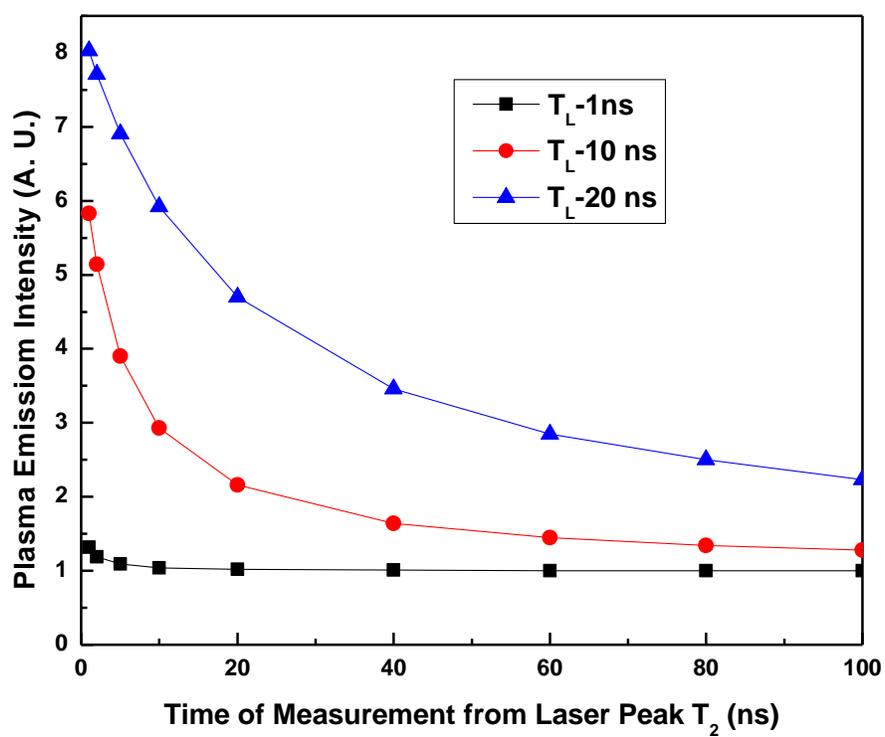

**Fig. -7**